\pdfoutput=1 

\documentclass[a4paper, 11pt, final]{article}

\usepackage{amssymb} % provides various useful mathematical symbols
\usepackage{amsthm} % provides extended theorem environments
\usepackage{amsmath,empheq}
\usepackage{bbm}

\DeclareMathAlphabet{\mathcal}{OMS}{cmsy}{m}{n}
\DeclareMathAlphabet\mathbfcal{OMS}{cmsy}{b}{n}

\DeclareFontFamily{U}{dutchcal}{\skewchar\font=45 }
\DeclareFontShape{U}{dutchcal}{m}{n}{<-> s*[1.0] dutchcal-r}{}
\DeclareFontShape{U}{dutchcal}{b}{n}{<-> s*[1.0] dutchcal-b}{}
\DeclareMathAlphabet{\mathcald}{U}{dutchcal}{m}{n}
\SetMathAlphabet{\mathcald}{bold}{U}{dutchcal}{b}{n}
\DeclareMathAlphabet\mathcalz{T1}{pzc}{mb}{it}

\usepackage[nice]{nicefrac} % nicer-looking fractions
\usepackage{siunitx} % standard units, e.g., scientific notation

\usepackage{newtxtext, newtxmath} % change font to Adobe Times New Roman

\usepackage{anyfontsize}
\usepackage[utf8]{inputenc}
\usepackage[T1]{fontenc}
\usepackage{microtype} % Slightly tweak font spacing for aesthetics

\providecommand{\JEL}[1]{\textit{\textbf{JEL: }} #1}
\providecommand{\keywords}[1]{\textbf{\textit{Keywords--- }} #1}

\usepackage{setspace} 

\usepackage{parskip} % each new line automatically spaces previously paragraph correctly
\usepackage{etoolbox}
\usepackage{titlesec}
\titleformat{\section}{\normalfont\Large\bfseries}{\thesection}{1em}{}
\titleformat{\subsection}{\normalfont\large\bfseries}{\thesubsection.}{1em}{}
\titleformat{\subsubsection}{\normalfont\normalsize\itshape}{\thesubsubsection.}{1em}{}

\usepackage{abstract}
\renewenvironment{abstract}
 {\normalfont
  \begin{center}
  \bfseries \abstractname\vspace{-.5em}\vspace{0pt}
  \end{center}
  \list{}{
    \setlength{\leftmargin}{0cm}%
    \setlength{\rightmargin}{\leftmargin}%
  }%
  \item\relax}
 {\endlist}

\usepackage{authblk} % package for author affiliations
\usepackage[bottom]{footmisc} % Makes footnotes stick to bottom of the page

\usepackage{multicol} % for multi-tab enumerated lists

\usepackage{enumitem} % for more flexibility in lists; see https://mirror.ufs.ac.za/ctan/macros/latex/contrib/enumitem/enumitem.pdf

\usepackage{graphicx} % More advanced figure inclusion
\usepackage{float} % For specifying table/figure locations, i.e. [ht!]
\usepackage{subcaption}
\usepackage{afterpage} % For encapsulating a floating figure on a single page   

\usepackage{printlen}

\usepackage[labelfont=bf]{caption}
\usepackage{color, colortbl}
\definecolor{LightGray}{rgb}{0.93,0.914,0.914}   
\usepackage{soul} % for highlighting stuff
\usepackage{longtable,rotating} % For long tables that span multiple pages
\usepackage{booktabs} % used for making more professional-looking tables
\usepackage{multirow} % used for cell-merging in complex tables
\usepackage{arydshln} % for drawing other line types in tables
\usepackage{bigdelim} % for curly braces; see https://tex.stackexchange.com/questions/164546/rotate-text-next-to-right-brace-in-multirow-environment
\usepackage{tabularx}
\usepackage{threeparttable}
\usepackage[table]{xcolor}

\usepackage[most]{tcolorbox}            % for boxed environments in colour

\PassOptionsToPackage{hyphens}{url} % for hyphenating URLs

\usepackage{fancyhdr}
\fancyheadoffset{0cm}
\makeatletter
\newcommand{\quickwordcount}[1]{
  \immediate\write18{texcount -quiet -incbib -sub=none -utf8 -1 -sum -merge -encoding=utf8 #1.tex > #1-words}%
  \immediate\openin\somefile=#1-words
  \read\somefile to \@@localdummy
  \immediate\closein\somefile
  \setcounter{wordcounter}{\@@localdummy}
  \@@localdummy
}
\makeatother

\usepackage{silence}
\usepackage[style=apa,backend=biber,natbib,hyperref]{biblatex}
\DeclareLanguageMapping{british}{british-apa}
\defbibenvironment{bibliography}{\enumerate}{\endenumerate}{\item}

\usepackage[colorlinks=true, allcolors=gray]{hyperref}
\let\orgautoref\autoref

\renewcommand{\autoref}[1]
{%
\def\equationautorefname{Eq.}%
\def\sectionautorefname{Sec.}%
\def\subsectionautorefname{Subsec.}%
\def\figureautorefname{Fig.}%
\def\subfigureautorefname{Fig.}%
\orgautoref{#1}%
}

\usepackage[nameinlink,capitalise]{cleveref}
\usepackage{algorithm,algpseudocode}

\makeatletter
\newlength{\trianglerightwidth}
\algnewcommand{\LineCommentCont}[1]{\Statex \hskip\ALG@thistlm%
  \parbox[t]{\dimexpr\linewidth-\ALG@thistlm}
{\leftskip=\algorithmicindent
  \hangindent=\algorithmicindent 
  \hangafter=1%
  \strut\makebox[\algorithmicindent][c]{$\triangleright$}#1\strut}
  } % \trianglerightwidth
\makeatother

\usepackage[left=1.8cm, right=1.8cm, bottom=2.5cm, top=2.5cm]{geometry}

\begin{document}

%TC:ignore

% more Hyperref options (must be after \begin{document}
\renewcommand{\figureautorefname}{Fig.}
\onehalfspacing

%--------------------------------------------------------%
%	TITLE PAGE
%--------------------------------------------------------%

%--------------------------------------------------------%
%	TITLE
%--------------------------------------------------------%

% Article title
\newcommand{\MainTitleText}{A cost of capital approach to determining the LGD discount rate}

% Title candidates:
% 1) A dynamic model for IFRS 9 stage impairment classification
% 2) Definitions for classifying staged impairment under IFRS 9
% 3) Defining SICR-events under IFRS 9: a dynamic model for classifying impaired loans
% 4) Defining SICR-events for classifying impaired loans under IFRS 9: A comparative study
% 4) Defining and comparing SICR-events for classifying impaired loans under IFRS 9

\title{\fontsize{20pt}{0pt}\selectfont\textbf{\MainTitleText
}}

%--------------------------------------------------------%
%	AUTHORS
%--------------------------------------------------------%   

\author[,a,c]{\large Janette Larney \thanks{ ORC iD: 0000-0003-0091-9917; Corresponding author: \url{janette.larney@nwu.ac.za}}}
\author[,a,c]{\large Arno Botha \thanks{ ORC iD: 0000-0002-1708-0153}}
\author[b]{\large Gerrit Lodewicus Grobler}
\author[a,c]{\large Helgard Raubenheimer}
\affil[a]{\footnotesize \textit{Centre for Business Mathematics and Informatics, North-West University, Private Bag X6001, Potchefstroom, 2520, South Africa}}
\affil[b]{\footnotesize \textit{Department of Statistics, North-West University, Private Bag X6001, Potchefstroom, 2520, South Africa}}
\renewcommand\Authands{, and }
\affil[c]{\footnotesize \textit{National Institute for Theoretical and Computational Sciences (NITheCS), Stellenbosch 7600, South Africa}}

% Today's date
 	%\date{Submitted: \usvardate\today}
    
%by specifying the below, we essentially "rewrite" the command \maketitle, which is normally called in main.tex.
%this is done primarily to abuse the \date command above

\makeatletter
\renewcommand{\@maketitle}{
    \newpage
     \null
     \vskip 1em%
     \begin{center}%
      {\LARGE \@title \par
      	\@author \par
        }
     \end{center}%
     \par
 } 
 \makeatother
 
 \maketitle

%--------------------------------------------------------%
%	ABSTRACT
%--------------------------------------------------------%    
{
    \setlength{\parindent}{0cm}
    \rule{1\columnwidth}{0.4pt}
    \begin{abstract}
        Loss Given Default (LGD) is a key risk parameter in determining a bank's regulatory capital. During LGD-estimation, realised recovery cash flows are to be discounted at an appropriate rate. Regulatory guidance mandates that this rate should allow for the time value of money, as well as include a risk premium that reflects the "undiversifiable risk" within these recoveries. Having extensively reviewed earlier methods of determining this rate, we propose a new approach that is inspired by the cost of capital approach from the Solvency II regulatory regime. Our method involves estimating a market-consistent price for a portfolio of defaulted loans, from which an associated discount rate may be inferred. We apply this method to mortgage and personal loans data from a large South African bank. The results reveal the main drivers of the discount rate to be the mean and variance of these recoveries, as well as the bank's cost of capital in excess of the risk-free rate. Our method therefore produces a discount rate that reflects both the undiversifiable risk of recovery recoveries and the time value of money, thereby satisfying regulatory requirements. This work can subsequently enhance the LGD-component within the modelling of both regulatory and economic capital.
    \end{abstract}
     
     % Insert keywords here
     
     % Insert JEL codes here
     \JEL{G22, G32, C54.}
    
    \rule{1\columnwidth}{0.4pt}
}

\noindent Word count (excluding front matter and appendix):  7374 %\quickwordcount{ms} 

\noindent Figure count: 5

%TC:endignore

%--------------------------------------------------------%
%	CONTENT
%--------------------------------------------------------%

\subsection*{Disclosure of interest and declaration of funding}
\noindent This work is not financially supported by any institution or grant, with no known conflicts of interest that may have influenced the outcome of this work. The authors would like to thank all anonymous referees and editors for their valuable contributions.

\newpage

\section{Introduction}
\label{sec:Intro}

\begin{figure}[ht!]
\centering\includegraphics[width=0.77\linewidth,height=0.25\textheight]{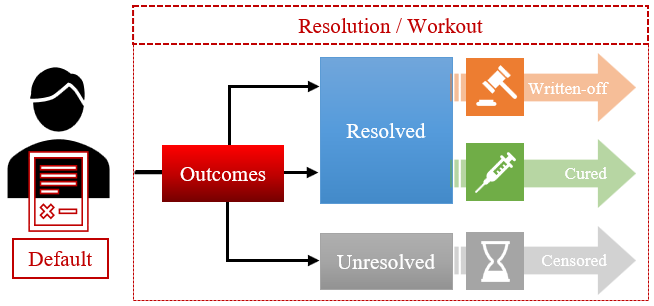}
\caption{Illustrating the typical process of resolving a defaulted loan into either a written-off or cured outcome during the workout period. The remaining defaults are considered as right-censored. From \citet[pp.~59]{botha2021Proc}.}\label{fig:DefaultOutcomes}
\end{figure}

% What is LGD?
One of the key risk parameters used in determining a bank’s regulatory capital under the Basel II Capital Accord from the \citet{basel2006} is the \textit{loss given default} (LGD). Denoted as the random variable $L\in[0,1]$, a loan's LGD is the fraction of its  balance $B_{\tau_d}$ (or the \textit{exposure at default}, EAD) that was eventually lost on a defaulted account, as at the time of default $\tau_d$. The LGD is often described as the "economic loss" on a financial instrument, which includes both material discount effects and material direct/indirect collection costs; see \citet[\S 460]{basel2006}, itself corroborated by the \citet{european2006directive}. 
% Meaning of successful (cured) collection and associated LGD-value
In nursing the strained relationship between the bank and the defaulted borrower back to health, collection staff can pursue a variety of remedial actions during this so-called \textit{resolution/workout period}. From \citet[pp.~26-28]{VanGestel2009book}, \citet[pp.~11-13]{finlay2010book}, and \citet[\S2.2]{botha2021Proc}, these remedial actions commonly include: 1) temporarily lowering instalments and/or zeroing interest rates; 2) extending a payment `holiday' during which instalments are suspended; and 3) restructuring the loan in the borrower's favour.
Successful collection efforts imply the distressed borrower repaying a newly-negotiated stream of cash flows over time and/or lump sums, both of which are called \textit{receipts} in this work. These receipts should diminish the arrears balance and decrease the delinquency, ultimately `curing' the loan back to health; resulting in a zero credit loss $L=0\%$. 
% Meaning of failed collection and associated LGD-value
Conversely, collection efforts may fail and further erode the credit relationship beyond repair. The bank will now initiate legal proceedings and/or foreclose on any available collateral, in recovering as much of the defaulted debt in record time. Any last receipts resulting from debt recovery are offset against the loan's last-known balance, with the non-zero remainder being written (or charged) off as a credit loss, i.e., $L>0\%$. Accordingly, a defaulted loan will resolve into either a \textit{cured} or a \textit{written-off} outcome, with the remaining unresolved cases being right-censored (pending an outcome). These ideas are illustrated in \autoref{fig:DefaultOutcomes}.

%although realisations $l$ can exceed these boundaries in extreme cases. For example, upon selling underlying loan collateral (if available), unexpected gains in the value thereof can exceed $B_\tau$ and yield negative $l$-values. Conversely, the combination of recovery costs and zero recoveries can produce cases with $l>100\%$. However, many regulators impose sensible boundaries on $l$, thereby circumventing the issue entirely; see \citet[\S 10]{baesens2016credit}.

\begin{figure}[ht!]
\centering\includegraphics[width=0.85\linewidth,height=0.34\textheight]{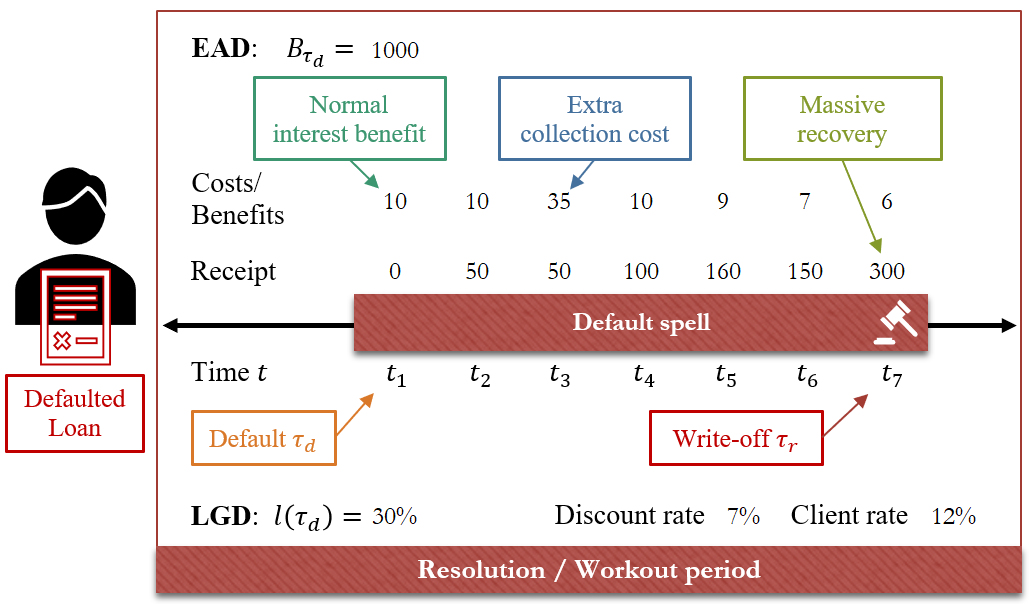}
\caption{An example of using the workout method to calculate a written-off loan's LGD-value from a bank's own credit data. Given a default balance of $B_{\tau_d}=1000$ and net cash flows (receipts minus costs/benefits) over subsequent times $t=\tau_d,\dots,\tau_r$, the associated LGD is calculated as $l(\tau_d)$ at default time $\tau_d$ using \autoref{eq:realised_loss}.}\label{fig:LGD_WorkoutMethod}
\end{figure}

% Estimation of LGD
Under the advanced \textit{internal rating based} (IRB) approach of Basel, a bank can produce its own LGD-estimates. From \citet{schuermann2004we}, \citet[pp.~217-222]{VanGestel2009book}, and \citet[\S 10]{baesens2016credit}, there are four common methods for estimating the LGD of either retail or corporate exposures.
Of these methods, we shall focus on the more popular \textit{workout} method, which caters to both retail and corporate exposures.
% Mathematical description of LGD's calculation
At each month-end time period $t$, let $X_{it}$ denote the net cash flow (or receipt minus all costs \& benefits) of account $i=1,2,...,n$ over applicable time periods $t=\tau_d(i),\dots,\tau_r(i)$ for loan $i$, as observed from data. Similarly, let $B_{it}$ record the month-end outstanding balance or utilised amount of a credit limit that was observed from data. 
For a written-off loan $i$, the realised loss $l_{i}(u)$ is then calculated at (and from) a given time $u$ as the proportion of $B_{iu}$ that was ultimately lost. This given time $u$ can range from the moment of default $\tau_d(i)$ of loan $i$, up until its resolution/write-off time $\tau_r(i)\geq\tau_d(i)$; see \citet{gurtler2013LGD}, \citet[\S10]{baesens2016credit}, and \citet{scheule2020benchmarking}. Using an appropriate factor $v_m$ that discounts a quantity $m$ periods backwards, the realised loss (given write-off) is expressed more formally as \begin{equation}\label{eq:realised_loss}
    %l_{i}(u) = \frac{ B_{iu} - \sum_{t=u}^{\tau_r(i)}{ X_{it} v_{t-u} } }{B_{iu}} 
    l_i(u) = 1 - \frac{1}{B_{iu}} \sum_{t=u}^{\tau_r(i)}{ X_{it} v_{t-u} } \quad \text{for} \ u= \tau_d(i),\dots, \tau_r(i) \ \text{and} \ B_{ij}>0 \, .
\end{equation}
% Illustrate workout method
In preparing the modelling dataset, \autoref{eq:realised_loss} is typically evaluated only once at the default time $\tau_d(i)$ of each resolved defaulted loan $i$ that was eventually written off. By implication, these $l_i\left(\tau_d\right)$-values are considered as loan-level realisations from $L$ across all written-off loans $i$. %in respect of the defaulted loan $i$. 
Finally, we illustrate the workout method in \autoref{fig:LGD_WorkoutMethod} for a single hypothetical loan that was written-off.

% Discuss discount rate and position as main problem for this paper
In \autoref{eq:realised_loss}, selecting the appropriate discount rate in $v_m$ can be non-trivial and even contentious in correctly accounting for both the time value of money and any inherent risks. Moreover, the cumulative effect of any discount rate on $l_{i}(\tau_d)$ becomes particularly pronounced for longer resolution periods, e.g., as in residential mortgages. \citet[pp.~220-221]{VanGestel2009book} noted that the most appropriate rate ought to be the "risk-free" rate $r_f$. 
Furthermore, and from a regulatory perspective, the \citet{basel2005LGD} offers principled (but limited) guidance on selecting an appropriate discount rate, which is arguably inspired by the seminal work of \citet{markowitz1952portfolio}. In short, discount rates must reflect the time value of money ($r_f$) as well as include an appropriate risk premium $\delta$ (for "undiversifiable risk"), which compensates for the uncertainty of receiving $X_{it}$. 
We therefore structure this so-called \textit{Basel-principle} as follows, whereupon we rewrite the discount factor $v_m$ in \autoref{eq:realised_loss} using the discount rate $r_d \in [0,1]$; i.e., \begin{equation} \label{eq:Discount_Factor}
    \text{Basel-principle:} \quad r_d=r_f+\delta \quad \implies \quad v_m=\frac{1}{(1+r_d)^{m}} \, .
\end{equation}

% Problem: How to determine risk premium? Variety of methods, promulgated by different regulators
Some regulators have adopted this principle \textit{verbatim} in selecting a discount rate, e.g., see the retail credit template of the South African regulator, the \citet[p.~20]{sarb2018long}. 
% EU-regulator
Others have digressed somewhat, e.g., the EU-regulator (European Banking Authority) proposed using a discount rate comprised of the primary interbank rate plus 5\%; see \citet{eba2017PDandLGD}. Though simple, this proposal deliberately ignores a bank's funding costs when assessing the uncertainty of receiving $X_{it}$; itself affected by competent collection staff and associated indirect costs.
% UK-regulator
Similarly, the UK-regulator (Prudential Regulation Authority [PRA]) suggests a discount rate that consists of the \textit{Sterling Overnight Index Average} (SONIA) as the interbank rate, plus 5\%; see \citet{pra2019PD_LGD} and \citet{pra2020PD_LGD}. In addition, the PRA imposes a minimum of 9\% for discount rates used when estimating downturn LGD-values.
% Implication
It is quite clear to us that there exists little consensus regarding even the prescription of $r_d$ amongst regulators, and perhaps even less amongst practitioners.

% Solution to problem: CoC-approach
As an alternative, we propose a \textit{cost of capital} (CoC) approach for inferring the risk premium $\delta$ based on a market-consistent (or fair) price for defaulted exposures. Inspired by the Solvency II (SII) regulatory regime for insurers \citep{solvency2009directive}, our CoC-approach can produce a $\delta$-value that reflects the undiversifiable risk within $X_{it}$. 
% Proximity of our work to most relevant works
Our work is closest to that of \citet{altman2003market}, \citet{brady2006discount}, \citet{witzany2009unexpected}, and \citet{jacobs2012empirical}, wherein the discount rate was similarly inferred from market prices of defaulted debt, which typically includes corporate bonds and traded bank loans. These studies rely on the \textit{“informational efficient market"} proposition of \citet{fama1970efficient}, which assumes that market prices incorporate all available information about future values. 
As such, the inferred discount rate should objectively reflect the market's confidence on receiving $X_{it}$ in future, provided that the market is sufficiently deep and liquid; see \citet{jankowitsch2014determinants}.
% But what differentiates our work from theirs?
However, these procedures fail in the absence of such market prices, which is often the case for retail loan portfolios. Our CoC-approach can generate such market-consistent (but synthetic) prices, driven by the cost of holding capital against uncertain cash flows.
%\textcolor{red}{AB: How does our CoC-approach solve for this?}
% synthetic market price
% S2 requires an 'economic' evaluation of balance sheet, i.e., A&L must be shown at market value. If market prices are unavailable, then generate market-consistent price, hence CoC-approach.
% Needs compensation (or risk capital) for holding these risky defaulted assets, but such capital has cost. delta => IRR of market-consistent price (which reflects cost of capital) of defaulted portfolio. Buyer pays E(X_{it}), but reduce with CoC against risky/volatile cashflows.
%%% AB: Our work is also close to Witzany, which also uses a CoC-approach, albeit calculated from regulatory capital. 

The remainder of this paper is organised as follows. In \autoref{sec:earlierMethods}, we provide an overview of the most relevant methods from literature in determining the LGD discount rate.
We outline in \autoref{sec:regulation_solvency} the fundamental ideas of the SII regulatory regime in calculating the cost of capital. 
Thereafter, the formulation of our CoC-approach is provided in \autoref{sec:Method} towards calculating a \textit{market-consistent price} (MCP) for a stream of uncertain cash flows $X_{it}$.
In \autoref{sec:Data&model}, we calibrate this CoC-approach using two different credit datasets from a large South African bank: a secured mortgage portfolio and an unsecured personal loans portfolio.
Finally, the CoC-approach is illustrated in \autoref{sec:Results}, thereby obtaining an MCP from which $\delta$ can be inferred, whereafter we conclude the study in \autoref{sec:Conclusion}.

%The application of the CoC-methodology to determine a MCP for a portfolio of defaulted loans is then described, and we show how a market-consistent discount rate can be inferred from this MCP.
%In \autoref{sec:Data&model} we describe the two data sets, each relating to a different retail portfolio of a large South African bank, used to illustrate our proposed CoC-methodology. The results hereof are reported in \autoref{sec:Results}, followed by concluding in \autoref{sec:Conclusion}.
%\textcolor{red}{AB: Link to codebase?}

\section{Earlier methods for determining the discount rate \texorpdfstring{$r_d$}{Lg}}
\label{sec:earlierMethods}

In determining an appropriate $v_m$-value for \autoref{eq:realised_loss}, a few approaches exist in literature. 
% Outline Scheule2020's 5 approaches, focusing on contract rate
\citet{scheule2020benchmarking} studied and compared a few different methods of calculating a discount rate for LGD-modelling, having focused on the robustness and empirical implications of these methods. These methods respectively use: 1) the loan's \textit{contract rate} at origination; 2) the loan's \textit{expected return} (ER); 3) the market's average \textit{return on defaulted debt} (RODD); 4) the bank's \textit{cost of equity} (or ROE); 5) the \textit{market equilibrium} (ME) variant of ROE using Basel asset correlations; and 6) the \textit{weighted average cost of capital} (WACC). 
% Explain basis of contract rate method
As the most popular, the first method simply uses the contract rate $r_i$ of loan $i$ when discounting the cash flows $X_i=\left\{X_{it_1}, X_{it_2}, \dots \right\}$ in \autoref{eq:realised_loss}. This method was first proposed by \citet{asarnow1995measuring} in measuring the loss on defaulted corporate bonds held by Citibank. 
In its simplest form, the contract rate $r_i\in[0,1]$ is constructed as the sum of a base rate $\rho_c$, a pre-selected profit margin $m$, and a factor $f_g$ for loans $i \in \mathcal{S}_g$ within each subset $S_g$ of the portfolio that coincides with the $g^{\text{th}}$ credit risk grade; i.e., the nominal annual rate $r_i=\rho_c+m+f_g$. A loan with greater credit risk therefore warrants a greater risk factor $f_g$, thereby increasing its `price' $r_i$; see \citet[\S3]{thomas2009consumer} and \citet{phillips2013optimizing} regarding risk-based pricing.

% Popularity and drawbacks of contract rate method
Given its availability at origination, the contract rate $r_i$ is a natural and rather popular choice for discounting cash flows $X_i$ when calculating realised LGD-values. In fact, \citet{scheule2020benchmarking} argued that its popularity is further entrenched by the two dominant accounting standards that guide impairment modelling: IFRS 9 from \citet{ifrs9_2014} and the \textit{Current Expected Credit Loss} (CECL) framework from \citet{cecl2016}. Both IFRS 9 and CECL require the use of the \textit{effective interest rate} (EIR), which is the rate that exactly discounts an impaired loan's future cash flows throughout its remaining expected life to the net carrying amount, i.e., the internal rate of return; see \S5.4 and \S B.5.4 of IFRS 9. Since future cash flows incorporate the $r_i$-based instalment, \citet{scheule2020benchmarking} noted that the EIR tends to be similar to $r_i$.
However, and despite its popularity, the authors note at least one major drawback of using $r_i$ in discounting $X_i$. Since $r_i$ embeds a loan's expected loss via $f_g$, the presumption is that the default event is uncertain (and hence unrealised) when pricing said loan at origination. However, calculating the realised LGD of loan $i$ is predicated upon first realising the default event, whereupon the subsequent cash flows $X_i$ are observed. This implies that using $r_i$ in discounting $X_i$ will overcompensate for the now-realised default risk, which can lead to artificially greater LGD-values. Doing so disagrees with the Basel-principle in that, as explained in both \citet{maclachlan2004choosing} and \citet{jacobs2012empirical}, the discount rate should only compensate for the uncertainty of receiving $X_i$ post-default, i.e., the so-called \textit{resolution risk}. Put differently, the contract rate exceeds the expected return when conditioning for default, as demonstrated by \citet{scheule2020benchmarking} in their second ER-method.

% Introduce RODD-method
Another simple approach for informing $r_d$ is the third so-called RODD-method, which is based on the price movements of tradable junk bonds. Given the price $\mathcal{B}_{it}$ of such a defaulted bond $i$ at time $t$, the realised return $\rho_i$ is expressed over the associated resolution period $\tau_r - \tau_d$ as 
\begin{equation} \label{eq:realisedReturn}
    \rho_i = \left(\frac{\mathcal{B}_{i\tau_r}}{\mathcal{B}_{i\tau_d}}\right)^{\frac{1}{\tau_r - \tau_d}} - 1 \, .
\end{equation}
The exact moment of default may not be known for a bond, though \citet{scheule2020benchmarking} remark that the bond price between 30 or 45 days past default is typically taken as a proxy for $\mathcal{B}_{i\tau_d}$. Having calculated the returns $\rho_i$ for a set of defaulted bonds $\mathcal{S}_b$, the expected return $\mathbb{E}(\rho_\mathrm{r})$ is estimated using the sample mean, which is then used as the value for the discount rate $r_d$ in $v_m$ within \autoref{eq:realised_loss}. The RODD-method is therefore summarised as 
\begin{equation} \label{eq:discountRate_RODD}
\text{RODD-method:} \quad \mathbb{E}\left( \rho_\mathrm{r}\right) \approx \frac{1}{\left\vert \mathcal{S}_b \right\vert}\sum_{i \, \in \, \mathcal{S}_b}{\rho_i} \quad \therefore \ r_d = \mathbb{E}\left( \rho_\mathrm{r}\right)  \, .    
\end{equation}
% Drawbacks of RODD-method
While certainly simplistic, the RODD-method can yield a negative value for $r_d$, which would be nonsensical in discounting the cash flows $X$ in \autoref{eq:realised_loss}. This can happen when the prices of most bonds within $\mathcal{S}_b$ have decreased at their respective resolution times $\tau_r$, i.e., $\mathcal{B}_{i\tau_r} < \mathcal{B}_{i\tau_d}$ holds for most $i$.

% ROE-method
Alternatively, the discount rate $r_d$ can be based more directly on the bank's funding costs by using the fourth ROE-method from \citet{scheule2020benchmarking}. Let $\rho_\mathrm{e}$ denote the returns over time for a hypothetical instrument that is comparable to a portfolio of defaulted loans, subject to resolution risk. Then, the expected return on equity $\mathbb{E}\left(\rho_\mathrm{e}\right)$ is estimated using the classical \textit{capital asset pricing model} (CAPM) from \citet{sharpe1964capital}, which is formalised as 
\begin{equation} \label{eq:discountRate_ROE}
    \text{ROE-method (CAPM):} \quad \mathbb{E}\left( \rho_\mathrm{e}\right) = r_f + \beta \cdot \underbrace{ \mathbb{E}\left( r_m - r_f \right)}_{\text{Market risk premium (MRP)}} \quad \therefore \ r_d = \mathbb{E}\left( \rho_\mathrm{e}\right)  \, ,
\end{equation}
where $r_f$ is the (known) risk-free rate and $r_m$ is the (unknown) market-related return of a similar portfolio.
% Candidates for r_m and r_f
In setting either parameter, \citet{scheule2020benchmarking} explained that  $r_m$ is often replaced with the equity returns on the local share market, e.g., the All-Share Index (ASI) from the \citet{jse2023} that tracks about 99\% of eligible securities. In fact, \citet{dimson2011equity} examined such equity returns relative to those of either government bonds or treasury bills (both of which can substitute $r_f$) and found that these "excess returns", or estimates of $\mathbb{E}\left( r_m - r_f \right)$, broadly averaged 3-8\% using 110 years of data across 19 countries.

% ROE-method's beta
The $\beta$ within the ROE-method from \autoref{eq:discountRate_ROE} measures the sensitivity of the instrument's returns to the variation of excess market returns. From \citet{fama2004capital} and \citet{scheule2020benchmarking}, this so-called \textit{market beta} is expressed as 
\begin{equation} \label{eq:discountRate_ROE_beta}
    \beta = \frac{\kappa\left(r_m, \rho_\mathrm{e} \right) \sigma_{\rho_\mathrm{e}} }{ \sigma_{r_m} } \, ,
\end{equation}
where $\kappa\left(X,Y\right)$ is the Pearson correlation between two random variables $X$ and $Y$, and $\sigma_{X}$ is the standard deviation of $X.$
% Candidates for beta
The authors noted previous studies that have used the NYU Bond Index and ninety defaulted US bonds as two candidates for $\rho_\mathrm{e}$, whereupon they obtained $\beta$-values between 35-40\%.

% "Equilibrium"-method (as variant on ROE)
\citet{maclachlan2004choosing} and \citet{scheule2020benchmarking} provided another interpretation of the correlation $\kappa\left(r_m, \rho_\mathrm{e} \right)$ in \autoref{eq:discountRate_ROE_beta} that leverages a method by which the regulatory capital of a bank is normally calculated. They use the \textit{asymptotic single-risk factor} (ASRF) model from \citet{vasicek2002distribution} and extended by \citet{gordy2003risk}; itself codified in the Basel framework from the \citet{basel2019}. 
In particular, the ASRF-model supposes that a corporate exposure $i$ defaults once its overall asset value $Y_i$ falls below a certain threshold. This $Y_i$-value is then structurally decomposed into: 1) a portfolio-level \textit{systematic} or market factor $V_\mathrm{m}$ that affects all exposures equally; and 2) an \textit{idiosyncratic} risk factor $Z_i$ specific to $i$. For the loan pair $i$ and $j$, the associated random variables $(Y_i,Y_j)$ are assumed to be bivariate normally distributed with standard normal marginals and with equal pairwise correlations $\kappa(Y_i, Y_j)=\kappa$, thereby accounting for correlated defaults when asset values move together. Finally, $Y_i$ is expressed using $\kappa$ with a Gaussian copula as $Y_i=\sqrt{\kappa}\cdot V_\mathrm{m} + \sqrt{1-\kappa}\cdot Z_i$; see \citet[\S2.5]{botha2021Proc}. 
The key economic insight from \citet{maclachlan2004choosing} is that $\sqrt{\kappa}$ behaves like $\kappa \left(r_m, \rho_\mathrm{e} \right)$ when substituting $\rho_\mathrm{e}$ with the market value of a defaulted exposure $i$. Accordingly, the authors provided a variant of the ROE-method (or the fifth ME-method) by redefining the $\beta$ within the CAPM from \autoref{eq:discountRate_ROE} as a \textit{defaulted debt beta}, expressed for firm $i$ as
\begin{equation} \label{eq:discountRate_ROE_beta_equilibrium}
    \text{ME-variant of ROE:} \quad \beta_i=\frac{\sqrt{\kappa} \cdot \sigma_i}{\sigma_{r_m}} \, ,
\end{equation}
where $\sigma_i$ denotes the standard deviation of the $i^{\text{th}}$ firm's asset value over time. The standard formulae for $\kappa$ can be used as prescribed by the Basel framework in \citet[\S CRE31]{basel2019}, e.g., a fixed 15\% for retail residential mortgages, or an exponentially-weighted interpolation between 12-24\% for corporate exposures based on their default risk. Accordingly, \citet{maclachlan2004choosing} illustrated this equilibrium-method with $\beta_i\in[0.4,1]$ across a range of regulatory asset classes, whereupon he obtained $r_d$-values of between 2.4-6\%.

% Drawbacks of CAPM towards outlining WACC-method
On its own, the CAPM has a few unrealistic assumptions that undermine its tractability in practice, which compromises the ROE-method. \citet{fama2004capital} examined some of these assumptions empirically and corroborated previous findings that ultimately rejected the CAPM. 
% Assumption: complete agreement
In particular, the CAPM assumes complete agreement amongst investors on the joint distribution of asset returns and its constituents, which assumes that $\mathbb{E}(\rho_\mathrm{e})$ from \autoref{eq:discountRate_ROE} is indeed linearly related to both the risk-free rate $r_f$ and the market beta $\beta$. The authors tested this assumption by regressing a time series of monthly portfolio returns on estimates of $\beta$, as calculated using \autoref{eq:discountRate_ROE_beta}. By implication, the intercept of such a simple regression model should approach $r_f$, while the coefficient on $\beta$ should approximate the excess market return $\mathbb{E}\left( r_m - r_f \right)$. While a positive linear relationship does exist between average returns and market betas, the real-world relationship is significantly `flatter' than predicted by the CAPM. In fact, these studies consistently find that the intercept is greater than the average $r_f$ (as proxied by the return on a 1-month US Treasury Bill), while the coefficient on $\beta$ is less than the average excess market return (as proxied by the return on common US stocks minus that of a 1-month Treasury Bill).
% Assumption: beta sufficiency
\citet{fama2004capital} also reviewed other studies in showing that $\beta$ is not the only factor that explains expected returns. Ratio-type variables like earnings-price, debt-equity (or leverage), and book-to-market can challenge the sufficiency of $\beta$ within the CAPM in predicting expected returns. This result is perhaps unsurprising since these variables embed information about the market's expectation of future returns, simply by incorporating the share price.
% Remedy: 3-factor Fama-French
As a viable alternative to the CAPM, the authors accordingly propose their \textit{3-factor Fama-French model}, which includes both firm size and book-to-market value as additional factors.

% WACC-method
While the ROE-method and its ME-variant may be alluringly simple, \citet{scheule2020benchmarking} noted that funding costs alone do not reflect the risk profile of subsequent lending activities, especially not the resolution risk thereof (cures/write-offs) given default.
Within the sixth so-called \underline{WACC-method}, the authors combine both the regulatory capital of a bank and its funding costs into a single rate, given that regulatory capital is a reasonable measure of systematic risk. 
% Formulate WACC-method
In particular, let $\rho_\mathrm{c}$ represent the return of an instrument that is comparable to a defaulted exposure subject to resolution risk. This return $\rho_\mathrm{c}$ is itself a blend of the costs of equity and debt-based funding, denoted respectively as $\rho_\mathrm{e}$ and $\rho_\mathrm{d}$. The CAPM from \autoref{eq:discountRate_ROE} can be used to estimate the expected equity return $\mathbb{E}\left(\rho_\mathrm{e}\right)$, while $\mathbb{E}\left(\rho_\mathrm{d}\right)$ can be derived from the weighted average cost of bank liabilities. The novelty lies in the so-called "funding weights" by which the estimates for $\mathbb{E}\left(\rho_\mathrm{e}\right)$ and $\mathbb{E}\left(\rho_\mathrm{d}\right)$ are eventually blended into the estimate for $\mathbb{E}\left(\rho_\mathrm{c}\right)$. Given default, the equity-based capital ratio $e_{\tau_d}$ may be loosely defined as the amount of capital held for the unexpected loss, divided by the post-default loan value (or expected recovery); i.e.,
\begin{equation} \label{eq:discountRate_WACC_CapitalRatio}
    e_{\tau_d} = \frac{ \big[\mathbb{E}\left(L | \text{Downturn} \right) -\mathbb{E}\left(L \right)\big] \cdot \text{EAD}}{ \big[1-\mathbb{E}\left(L \right) \big] \cdot \text{EAD}}  = \frac{\text{Bank Capital}}{\text{Remaining Loan Value}}\, ,
\end{equation}
where $\mathbb{E}\left(L | \text{Downturn} \right)$ is the expected LGD during an economic downturn.
Finally, the loan-specific $e_{\tau_d}$-estimate is used as the weight for blending the bank-level funding costs, summarised as 
\begin{equation} \label{eq:discountRate_WACC}
    \text{WACC-method:} \quad \mathbb{E}\left( \rho_\mathrm{c}\right) = e_{\tau_d}\mathbb{E}\left(\rho_\mathrm{e}\right) + \left(1-e_{\tau_d} \right)\mathbb{E}\left(\rho_\mathrm{d}\right) \, ,
\end{equation}
whereupon $r_d$ is set to the estimate of $\mathbb{E}\left( \rho_\mathrm{c}\right)$ and used in $v_m$ within \autoref{eq:realised_loss}. Naturally, the ROE-method is simply a special case of the WACC-method by setting the capital ratio $e_{\tau_d}=100\%$ in \autoref{eq:discountRate_WACC}, whereupon $\mathbb{E}(\rho_c)=\mathbb{E}(\rho_e)$.

% High-level comparison of methods from Scheule2020
\citet{scheule2020benchmarking} compared the aforementioned methods both qualitatively and quantitatively using global credit data across different geographies. 
The ROE-method from \autoref{eq:discountRate_ROE} often yielded the highest mean discount rate $r_d$ of 7.07\% with a resulting mean realised LGD $l(\tau_d)$ of 24.95\%. 
The contract rate method gave a significantly lower mean $r_d$ of 4.91\%, coupled with a mean $l(\tau_d)$ of 23.82\%. 
The ME-variant from \autoref{eq:discountRate_ROE_beta_equilibrium} tendered a lower mean $r_d$ of 4.13\% that yielded a mean $l(\tau_d)$ of 23.35\%.
As a variant of the contract rate method, the ER-method produced a lower mean $r_d$ of 3.72\%, which resulted in a mean $l_i(\tau_d)$ of 23.16\%
The WACC-method yielded an even lower mean $r_d$ of 3.26\%, which produced a mean $l(\tau_d)$ of 22.89\%. 
Lastly, and as a baseline, simply using the risk-free rate $r_f$ at default time $\tau_d$ gave the lowest mean $r_d$ of 2.36\% and a resulting mean $l(\tau_d)$ of 22.41\%.
% Implication
The last method's low rank might very well suggest that the resulting mean LGD might be underestimated.

% Witzany2009
Another notable method for estimating $r_d$ is the market-sensitive \textit{cost of risk capital} approach from 
\citet{witzany2009unexpected}, which might reflect the uncertainty of receiving cash flows. In particular, the author developed an iterative approach that is based on the CAPM as departure point, whilst blending in ideas from the ASRF-model under the Basel framework. This approach presupposes that realised loss rates can themselves be decomposed using a Gaussian copula between a systematic component and an idiosyncratic component, both of which are assumed to be independent standard normally distributed variables. Having calculated the expectation hereof using numerical procedures, the author provided an expression for the unexpected loss based on the Value-at-Risk (VaR) at some probability level for market risk. This quantity then informs the LGD risk capital, which is multiplied by the market-implied cost of holding such capital towards obtaining an associated $r_d$. In implementing this approach, a starting guess for $r_d$ is needed, whereupon the realised loss rates are obtained and the rest of the method is calibrated accordingly; which ultimately produces a new $r_d$-value. This iterative procedure continues until the difference between subsequent $r_d$-values becomes suitably small.
% Implication
Our method shall borrow elements hereof in determining $r_d$ using the market-sensitive cost of capital.

% Economic capital
In determining a suitable risk premium $\delta$, only a few other studies have investigated the relation between the MRP under the CAPM from \autoref{eq:discountRate_ROE} and the "cost of risk", or \textit{economic capital}. \citet{maclachlan2004choosing}, who developed the aforementioned ME-variant of the ROE-method, described economic capital as a risk charge held for the difference between a portfolio's maximum diminution in market value and its expected value. Even for defaulted exposures, it is deemed appropriate to hold economic capital in covering the systematic uncertainty within recovery cash flows; see \citet{frye2000collateral},  \citet{BCBS2004}, \citet{dullmann2004systematic} and \citet{caselli2008sensitivity}.
In extending this ME-method from \citet{maclachlan2004choosing}, the authors \citet{gibilaro2007selection} tweaked its reliance on the single risk factor from the ASRF-model, and replaced it with a multi-factor model that can better explain the variance in the LGD.
But to the best of our knowledge, a closed-form expression for economic capital is not widely known, except for the more esoteric proposals from \citet{tasche2004} and \citet{weissbach2010economic} using numerical procedures.
% Tasch2004
In particular, \citet{tasche2004} proposed using a variant of the ASRF-model wherein the LGD is treated as a random variable towards incorporating PD-LGD correlation. 
As mentioned earlier, in the ASRF-model defaults are driven by a random variable $Y$, called a risk-factor, that represents the overall asset value. In the Tasche-model, $Y$ not only drives the default event, but also the size of the loss incurred. LGD is then expressed as $L=\mathbb{I}(Y\leq \xi)G(Y),$ where $\xi$ is a chosen threshold value and $G$ is an increasing or decreasing function. Expressing LGD as a function of the default event in this way was first proposed in a two-factor model by \citet{pykhtin2003recovery}, where the default event and the loss were driven by two random variables.
However, the Tasche-model does not require estimating additional parameters and fits rather comfortably within the mathematical machinery of the existing Basel-framework.
Given its simplicity and ease of implementation, we shall use the Tasche-model in estimating economic capital as an input within our CoC-approach.

\section{The cost of capital under the Solvency II (SII) regime}
\label{sec:regulation_solvency}

SII is a regulatory regime aimed at insurance and reinsurance undertakings in the European Union (EU), as enacted by the \citet{solvency2009directive}. Several non-EU countries, including South Africa\footnote{Solvency Assessment and Management (SAM) represents a “Solvency II equivalent” risk-based supervisory approach for the prudential regulation of South African long-term and short-term insurers. See \url{https://tinyurl.com/tawdn8mz}}, Australia and Singapore have adopted SII-equivalent risk-based capital regimes, as discussed by  \citet{ferguson2009solvency}. Similar to the Basel framework for banks, SII is based on the principle that institutions should hold enough capital to cover the potential (unexpected) future losses that could result from their risk-inducing activities. However, SII and Basel differs in setting the required capital level, whilst differing also in the specific rules and regulations that they impose.

One of the key components of SII is the so-called "economic balance sheet" approach, where capital requirements are based on a balance sheet that reflects the economic value of a firm's assets and liabilities. Both assets and liabilities must be valued at \textit{fair value}, i.e., "the price that would be received to sell an asset or paid to transfer a liability in an orderly transaction between market participants at the measurement date", as defined by \citet{ifrs13_2011}. In most cases, the fair value is simply the current market value of an asset or liability, or alternatively, the market value of a replicating portfolio \citep{boekel2009replicating}.
If such a market value does not exist (i.e., the specific market is not deep, liquid, or transparent), then SII requires valuation to be at least market-consistent. To achieve such consistency, the "best estimate plus risk margin" approach has been adopted, where the \textit{best estimate of the present value} (or BEPV) of future cash flows is adjusted by a \textit{risk margin} (RM), or the market value margin. In turn, this particular BEPV is defined by SII as the "probability-weighted average of future cash flows", which incorporates the time value of money via discounting at least at the risk-free rate.

The idea with the RM is that a potential buyer of liabilities should be compensated for the cost of holding risk capital (economic or regulatory) in excess of the BEPV of the associated cash flows. Doing so will protect this buyer against the prospect of the BEPV-value being too low.
Similarly, an asset buyer should be compensated for the uncertainty surrounding the BEPV of the asset’s future cash inflows. This RM therefore compensates the providers of capital, whose funds are used to absorb the risk associated with uncertain cash flows. As such, the RM is then subtracted from the BEPV of assets (or added to the BEPV of liabilities) to determine the \textit{market consistent price} (MCP) of the underlying cash flows. Regarding its calculation, the RM is expressed as the cost of holding the required amount of capital over the run-off period of the risky cash flows. The base principle hereof is that the price of an asset or liability should reflect the cost of holding risk capital against inherent risks over its lifetime. This is the so-called \textit{cost of capital} (CoC) approach from the SII regime, which is further discussed by \citet{pelkiewicz2020review}.

In determining an MCP for a portfolio of defaulted debt, one needs to calculate the price that a potential investor would be willing to pay for the risky cash flows, or recoveries. As per the CoC-approach, these cash flows that represent the recoveries expected from the defaulted loans are first discounted at the risk-free rate $r_f$. From the resulting BEPV, the cost of holding risk capital (against the risk of recovering less than the expected amount) is then deducted. Any such capital required at any future time $t$ is also discounted at the $r_f$ back to the transaction date. Doing so reflects the fact that the capital buffer will be invested in assets that earn at least at $r_f$. In calculating the RM, the CoC-approach then requires an estimate of the required capital level as well as its cost (in the form of a rate). A few methods exist for determining this capital level, e.g., the Tasche-model for determining economic capital from \citet{tasche2004}. More importantly, and regarding the capital cost rate, consider the net cost of holding risk capital, which is the marginal cost of raising (and holding) risk capital, minus the interest that can be earned on such capital at $r_f$. In this study, we consider a bank's own cost of equity in excess of $r_f$ as a capital cost rate, which represents a realistic view of the associated cost of raising and holding the necessary risk capital. Such a rate could be prescribed by the regulator, in a similar way that SII prescribes an CoC-rate of 6\% for insurance and reinsurance undertakings.

\section{Towards a market-consistent 
 risk premium \texorpdfstring{$\delta$}{Lg} }
 \label{sec:Method}

In \autoref{sec:method_RiskPremium_Algo}, we provide a numeric algorithm for finding a plausible $\delta$-value based on the market-consistent price (MCP). This algorithm accepts estimates of the overall cost of capital over time, for which we provide an illustration in \autoref{sec:method_tasche} using the work of \citet{tasche2004}.

% Given realisations $l_i\left(\tau_d\right), i=1,2,...,n$ from $L$, where $\tau_d$ is the time of default, the expected LGD, $\mathbb{E}[L]$, can be estimated by
% \begin{equation} \label{eq:bar_lgd}
%     \bar{L} = \frac{1}{n}\sum_i^n {l_i\left(\tau_d\right)}
% \end{equation}
% and the unbiased estimator of the LGD variance, $\mathbb{V}[L]$, by
% \begin{equation}\label{eq:var_lgd}
% \hat{\sigma}^2=\frac{1}{n-1}\sum_{i=1}^n \left(l_i\left(\tau_d\right)-\bar{L}\right)^2.
% \end{equation} 

\subsection{Determining \texorpdfstring{$\delta$}{Lg} given economic capital estimates: a generic algorithm}
\label{sec:method_RiskPremium_Algo}

Our aim is to determine the MCP for a portfolio of defaulted debt, and from this, calculate the associated risk premium $\delta$ for use within \autoref{eq:Discount_Factor}.
This process requires an appropriate amount of economic capital $C_t$ over discrete time $t=1,2,\dots$, as discussed in \autoref{sec:earlierMethods}. However, the economic capital (EC) is generally a function of the workout LGD, which in turn relies on the discount rate $r_d=r_f+\delta$; i.e., $C_t$ is at least a function of $\delta$. 
Accordingly, and using a given capital vector $\boldsymbol{C}=C_t(\delta)$ over $t$, we define the \textit{risk margin} $R$ for a portfolio of defaulted loans as the function
\begin{equation}\label{eq:RM}
    R\left(\boldsymbol{C},\delta \right) = \sum_{t=1}^{\tau_A}\frac{c \  C_{t}(\delta)}{\left(1+r_f\right)^{t}} \, ,
\end{equation}
where $c\in[0,1]$ is an appropriate cost of capital rate, and $\tau_A$ is the maximum time spent in default across the portfolio, i.e., $\tau_A=\max\limits_{1\leq i\leq n} \left\lbrace\tau_{r}(i)-\tau_{d}(i)\right\rbrace$.
Let $Y$ denote the present value of the portfolio-level recoveries $X_t=\sum_{i=1}^n X_{it}$ for loans $i=1,\dots,n$ at each period $t$, where $r_d = r_f+\delta$ is used as the discount factor. Since $r_f$ is treated as a constant, we express the so-called \textit{discounted recoveries} $Y$ as the function
\begin{equation} \label{eq:disc_rec}
    Y(\delta)=\sum_{t=1}^{\tau_A}\frac{X_{t}}{\left(1+r_f+\delta\right)^{t}},
\end{equation}

Equipped with these two components \crefrange{eq:RM}{eq:disc_rec}, our objective is to find the value of $\delta\geq0$ that minimises the squared difference between the MCP and the $Y(\delta)$. More formally, our objective function $f$ is 
\begin{equation} \label{eq:objF}
    f(\delta, R, Y)= \Big( 
  \underbrace{\left[Y(0)-R\big(\boldsymbol{C}, \delta\big) \right]}_{\text{ MCP}} - Y(\delta)\Big)^2 \, ,
\end{equation}
and we are interested in finding 
\begin{equation} \label{eq:delta}
    \delta_*=\mathop{\arg \min}\limits_{\delta\geq 0}{f(\delta, R, Y)} \, .    
\end{equation}
However, note that $C_t(\delta)$ is already a function of $\delta$, which is now itself a function of $C_t(\delta)$ given \autoref{eq:delta}; a circular reference. We deliberately break this circularity by pre-calculating $R$ using a preset (but amendable) value of $\delta$ towards calculating $\boldsymbol{C}$ within a broader iterative approach. Doing so allows the $Y(\delta)$-term within \autoref{eq:objF} to be varied when finding $\delta$ in \autoref{eq:delta}, whereupon a new value for $R$ can be calculated in restarting the process. 
Accordingly, we provide an iterative procedure in \autoref{alg:delta_algo} by which \autoref{eq:delta} can be implemented in finding a $\delta$-value. This algorithm has the following general steps:
\begin{enumerate}
    \item Set an initial value for $\delta=\delta_0$ and calculate the economic capital $\boldsymbol{C}^*=C_t(\delta_0)$ across all $t$ accordingly.
    \item Calculate the risk margin $R^*=R(\boldsymbol{C}, \delta_0)$ from \autoref{eq:RM} using $\boldsymbol{C}=\boldsymbol{C}^*$ and $\delta=\delta_0$.
    \item Calculate a baseline discounted recoveries $Y_0=Y(\delta)$ from \autoref{eq:disc_rec} for $\delta=0$.
    \item Find $\delta_i$ by minimising $f$ from \autoref{eq:objF} but setting $R=R^*$. I.e., 
    \begin{equation} \label{eq:delta_noncircular}
        \delta_i= \mathop{\arg \min}_{\delta\geq 0} \left\lbrace \left[ Y_0- R^* \right]-Y(\delta)\right\rbrace^2 \, .
    \end{equation}
    \item Update $\delta=\delta_i$ and repeat steps 1 to 4 until $\delta_i$ converges, given a preset convergence criterion.  
\end{enumerate}

% Algorithm for applying simple clustered sampling
\begin{algorithm}[ht!]
  \caption{Pseudocode for finding a $\delta$-value via optimisation of \autoref{eq:delta}.}\label{alg:delta_algo}
     \textbf{Input: $c, \tau_A, r_f$} \Comment{Cost of capital $c$; maximum time spent in default $\tau_A$; risk-free rate $r_f$} \\
    \textbf{Input: $X_t$} \Comment{Recovery cash flows $X_t$ over discrete time $t=1,\dots,\tau_A$} \\
    \textbf{Input: $\tilde{C}(a,b)$} \Comment{Function for calculating EC-vector $\boldsymbol{C}=C_t(a)$ using discount rate $a$ across $t=1,\dots,b$} \\
    \textbf{Output: $\delta^*$} \Comment{Outputs an estimate $\delta^*$ that minimises \autoref{eq:delta}}
\begin{algorithmic}[1]
    
    \State $\epsilon \gets \num{e-4}$ \Comment{Tolerance level for small values}
    \State $\phi^* \gets \num{e4}$ \Comment{Initial error value}
    \State $\delta^*\gets \delta_0$ \Comment{Initial value $\delta_0$ towards finding $\delta^*$ iteratively}
    \State $Y_0 \gets \tilde{Y}\left(\delta, \tau_A, r_f \right)$ \Comment{Calculate \autoref{eq:disc_rec} for $\delta=0$, implemented here as $\tilde{Y}$ with other inputs $\tau_A$ and $r_f$}
    \While {$\phi* > \epsilon$} \Comment{Convergence criterion: loop until the error $\phi*$ becomes suitably small}
        \State $C^* \gets \tilde{C}\left(\delta^*, \tau_A\right)$ \Comment{Estimate EC-vector for $\delta^*$, implemented here as $\tilde{C}$ with other input $\tau_A$}
        \State $R^* \gets \tilde{R}\left(C^*, c, \tau_A, r_f \right)$ \Comment{Calculate \autoref{eq:RM} for $C^*$, implemented here as $\tilde{R}$ with other inputs $c$, $\tau_A$, and $r_f$ }
        \State{ $f := \text{function}\big( \left[ Y_0 - R^* \right] - \tilde{Y}\left(\delta^*, \tau_A, r_f \right) \big)^2$} \Comment{Initiate $f$ from \autoref{eq:objF} using $R^*$ and $Y_0$ as `static' inputs}
        \State $\delta_i \gets \text{minimise}\left(f, \delta^*\geq0 \right)$ \Comment{Implement \autoref{eq:delta}, i.e., find $\delta^*\geq0$ that minimises $f$}
        \State $\phi^* \gets \vert \delta^* - \delta_i \vert$ \Comment{Update convergence criterion}
        \State $\delta^* \gets \delta_i$ \Comment{Update $\delta^*$}
    \EndWhile
\end{algorithmic}
\end{algorithm}

Having applied \autoref{alg:delta_algo} in calculating $\delta$, it so happens that $r_f + \delta$ is 
the internal rate of return when equating MCP to the discounted cash flows $X_t$. This represents the fair return offered by investing in the portfolio of defaulted loans with $\delta$ serving as compensation for the uncertainty associated with the income stream. This $\delta$-quantity is itself driven by the cost of raising and holding the required EC-amount in respect of the uncertain cash flows. 
As such, we argue that this risk premium $\delta$ is appropriate for inclusion in the LGD discount rate $r_d$ from \autoref{eq:Discount_Factor}.

\subsection{Estimating economic capital (EC) using the single-factor Tasche-model}
\label{sec:method_tasche}

According to Basel-framework, the EC-amount for covering total loss $L^{\prime}$ is calculated as
\begin{equation}\label{eq:EC}
\mathbb{E}\left[L^{\prime}|V_m=\Phi^{-1}(\alpha)\right]-\mathbb{E}[L^{\prime}],
\end{equation}
where $\alpha$ is chosen at a suitably high level, and $\Phi^{-1}$ is the inverse cumulative distribution function (c.d.f.) of a standard normal variate $Z$. 
Put differently, the EC-amount is calculated as the difference between the Value-at-Risk (VaR) of the loss distribution and the expected loss of the loan portfolio, as discussed by \citet[pp.~54--55]{botha2021Proc}. Then, and as outlined in Section \ref{sec:earlierMethods},  \citet{tasche2004} proposed a method that incorporates the PD-LGD dependency by expressing total loss $L^{\prime}$ as 
$$
L^{\prime}=\mathbb{I}\left( W \geq \Phi^{-1}(1-p)\right) F_D^{-1}\left(\frac{\Phi\left(W\right)-1+p}{p}\right),
$$
where $W = \sqrt{\kappa}V_m+\sqrt{1-\kappa}Z$ is the loss, $\kappa$ is a correlation coefficient, $F_D$ is the c.d.f. of the loss distribution, and $p$ is the probability of default.

While the expectation $\mathbb{E}[L^{\prime}]$ in \eqref{eq:EC} is simply calculated as $p\mathbb{E}[L^{\prime}|L^{\prime}>0],$ the conditional loss term is expressed as
\begin{equation} \label{eq:condiLoss}
    \mathbb{E}\left[L^{\prime}|V_m=\Phi^{-1}(\alpha)\right]=\int_{a}^{\infty}\varphi(z)F^{-1}_D\left(\frac{\Phi\left(\sqrt{\kappa}\Phi^{-1}(\alpha)+\sqrt{1-\kappa}z\right)-1+p}{p}\right)\mathrm{d}z \, ,
\end{equation}
where 
$$
a=\frac{\Phi^{-1}(1-p)-\sqrt{\kappa}\Phi^{-1}(\alpha)}{\sqrt{1-\kappa}}
$$
and $\varphi$ is the density of a standard normal variate.
By using Gauss quadrature numerical integration, \citet{tasche2004} proposes approximating the integral in \autoref{eq:condiLoss} with
\begin{equation} \label{eq:condiLoss_approx}
    \mathbb{E}\left[L^{\prime}|V_m=\Phi^{-1}(\alpha)\right]\approx \frac{p}{2\sqrt{1-\kappa}}\sum_{i=1}^5 w_iH(t_i,\Phi^{-1}(\alpha)),
\end{equation}
where
\begin{equation}
    H(t,\Phi^{-1}(\alpha))=\frac{\varphi\left(\frac{\Phi^{-1}(p(t+1)/2+1-p)-\sqrt{\kappa}\Phi^{-1}(\alpha)}{\sqrt{1-\kappa}}\right)}{\varphi\left(\Phi^{-1}(p(t+1)/2+1-p\right)}F^{-1}_D\left(\frac{t+1}{2}\right) \, , \nonumber
\end{equation}
and the weights $w_i$ and nodes $t_i$ are given by
\begin{align*}
t1&=-0.9061798459 &w_1&=0.2369268851 &\\
t2&=-0.5384693101 &w_2&=0.4786286705 &\\
t3&=0 &w_3&=128/225 &\\
t4&=-t2 &w_4&=0.2369268851 &\\
t5&=-t1 &w_5&=0.2369268851 \, . &
\end{align*}
% Brief introduction of Tasche-model
%From \citet{tasche2004}, 
%\textcolor{red}{GG to describe Tasche-model, focussing on implementation side}
%a parametric beta distribution is used to represent the random variable $L$ as the LGD. The expectation and variance of $L$ are respectively given by
%\begin{equation} \label{eq:LGD_exp_var}
%\mathbb{E}[L] =  \frac{a}{a+b} \quad \text{and} \quad \mathbb{V}[L] = \frac{ab}{(a+b)^2(a+b+1)} \, ,
%\end{equation}
%where $a$ and $b$ are the shape parameters of the beta distribution. 
%\textcolor{red}{AB: The following just describes the beta distribution.}

In calculating the EC using the Tasche-model, let the loss distribution $F_D$ be a beta distribution with shape parameters $a$ and $b$, which aligns with \citet{tasche2004}. Despite its flexibility, some studies have demonstrated the beta distribution to be inappropriate since it cannot readily capture the bimodal-shape of the LGD distribution; see \citet{Calabrese2010BankLR}, \citet{huang2012modeling} and \citet{larney2022}. However, and despite its shortcomings, the beta distribution remains a popular and rather practical choice given its ability to assume many distributional shapes in modelling the LGD parametrically.
It is also fairly straightforward to estimate its two parameters $a$ and $b$ by ways of moment matching, provided with the sample mean and sample variance of realised loss data.
%Given realisations $l_i = l_i\left(\tau_d\right)$ from $L$ across defaulted loans $i=1,2,...,n$, we estimate both $\mathbb{E}[L]$ and $\mathbb{V}[L]$ accordingly.
%\begin{equation} \label{eq:LGD_exp_var_EST}
 %   \bar{L} = \frac{1}{n}\sum_i^n l_i \quad \text{and} \quad \hat{\sigma}^2=\frac{1}{n-1}\sum_{i=1}^n \left(l_i-\bar{L}\right)^2 \,.
%\end{equation}
Finally, we set $C_t(\delta)$ equal to this EC-amount, which is determined at the start of each of the run-off years for a loan portfolio at the 99.9\% confidence interval. Doing so agrees with Basel requirements, as outlined by the \citet[pp.~11-12]{BCBS2003}. 
While we have used the Tasche-model in determining EC, a bank should consider its own strategic objectives and risk appetite when deciding the appropriate risk measure, confidence level, and time horizon to use.
 
\section{A description and analysis of real-world loss data}
\label{sec:Data&model}

We illustrate our CoC-approach by applying it to two retail credit datasets provided by a large South African bank. The first dataset comprises recovery data of an unsecured \textit{personal loans} (PL) portfolio, whereas the second dataset contains secured residential \textit{mortgage loans} (ML).
Each dataset includes cash flows (both positive and negative) received on accounts that defaulted during Oct-2005 to Sep-2011. This six-year period also includes an economic downturn period of two years, during which each portfolio suffered the highest default rates. In particular, the downturn periods respective to the PL- and ML-portfolios are \{Dec-2007 : Nov-2009\} and \{Sep-2007 : Aug-2009\}, both of which coincide approximately with the Global Financial Crisis of 2007; see \citet{olbrys2021global}.
In estimating the risk-free rate $r_f$, we first interpolate between the two closest available tenors on the South African bond yield curve and then average the interpolated rates for all trading days within the relevant reference period. For example, the mean workout period for the full period on personal loans is $4.67$ months. Therefore, for each day within our observed period, a risk-free rate is obtained by interpolating between the rates observed at a 3-month and 6-month tenor on the yield curve. The estimated risk-free rate of $6.37\%$ is then obtained as the arithmetic average of these daily rates over the observed period.

\begin{table}[hbt!]
\centering
\begin{threeparttable}
\caption{Summary statistics for PL- and ML- portfolios over the full and downturn periods} \label{tab:summarystats}
%\label{table_example}
\begin{tabular}{lcccccc}
\toprule
& $n$ & $\bar{L}$ & $\hat{\sigma}$ & Max workout period & Mean workout period & $r_f$ \\
\midrule
\multicolumn{7}{c}{Personal Loans} \\
\midrule
Full & $121,151$ & $0.749$ & $0.290$ & $87$ months & $4.67$ months & $6.37\%$\\
Downturn & $60,032$ & $0.760$ & $0.273$ & $80$  months & $4.57$  months & $8.93\%$\\
\midrule
\multicolumn{7}{c}{Mortgage Loans}\\
\midrule
Full & $126,580$ & $0.256$ & $0.366$ & $113$  months & $25.64$  months & $6.88\%$\\
Downturn & $60,779$ & $0.293$ & $0.372$ & $90$  months & $33.59$  months & $8.49\%$\\
\bottomrule
\end{tabular}
\end{threeparttable}
\end{table}

\begin{figure}[H]
  \centering
  \includegraphics[width=.9\linewidth]{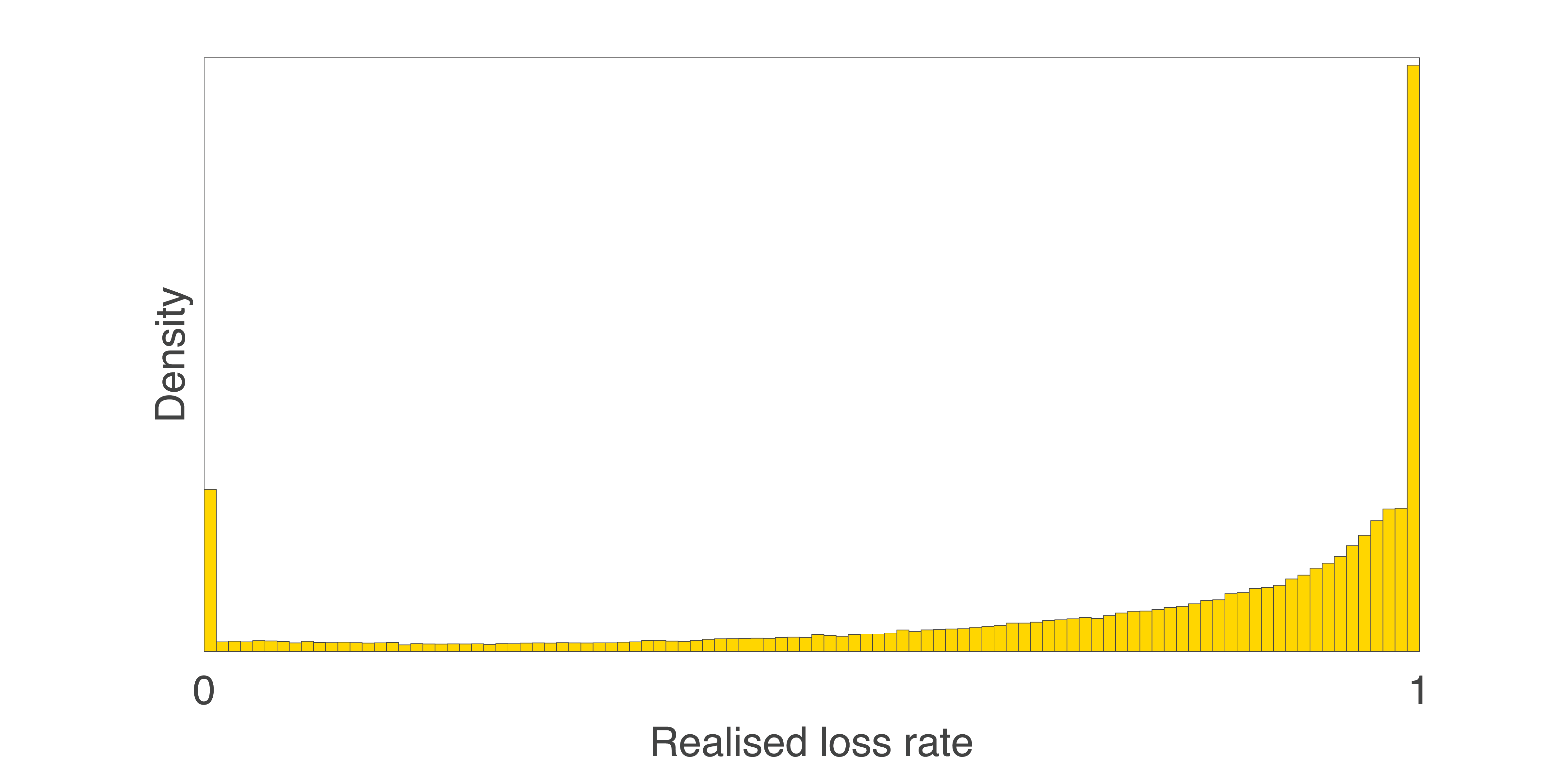}% Background/main image
  \makebox[-70pt][r]{% Similar to \llap
    \raisebox{5em}{%
      \includegraphics[width=0.6\linewidth]{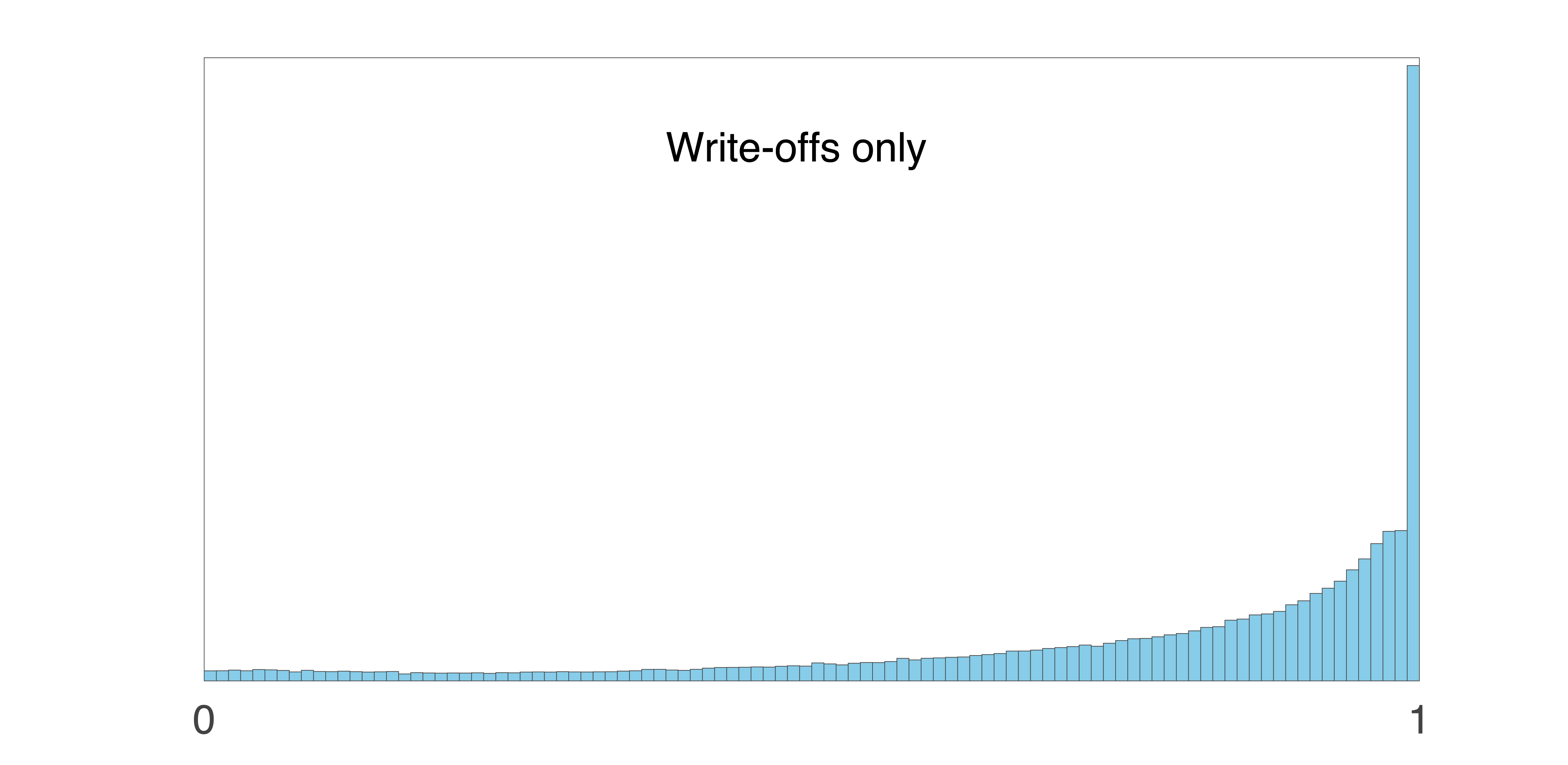}% Inserted image/inset
    }\hspace*{1em}%
  }%
  \caption{Histogram of realised loss rates for the unsecured PL-portfolio over the full reference period. The outer graph shows all data (including `cures'), whilst the inset graph depicts the distribution of those loans that were written-off. }
\label{fig:PLOutcomes}
\end{figure}

\begin{figure}[H]
  \centering
  \includegraphics[width=.9\linewidth]{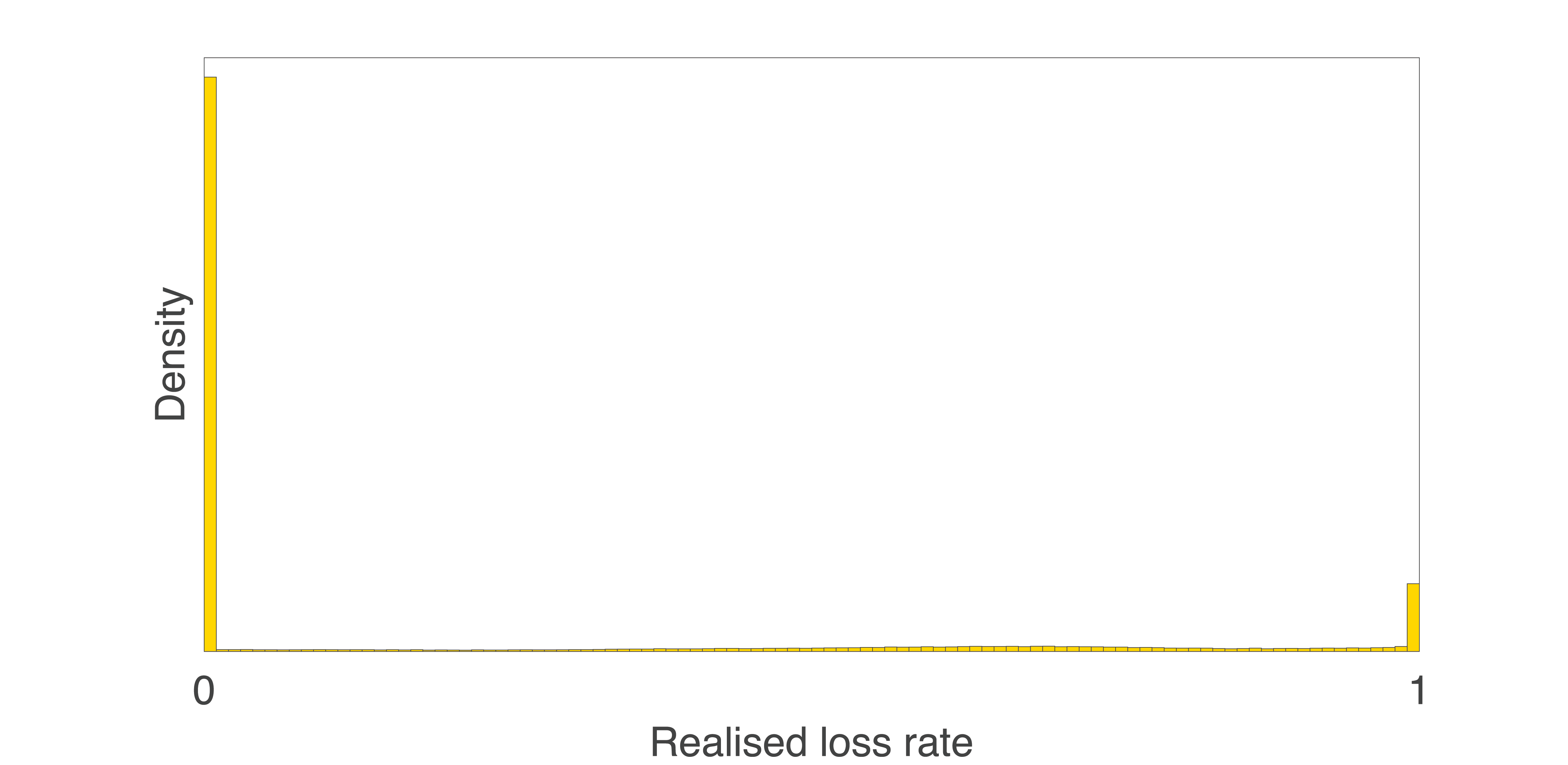}% Background/main image
  \makebox[-70pt][r]{% Similar to \llap
    \raisebox{5em}{%
      \includegraphics[width=0.6\linewidth]{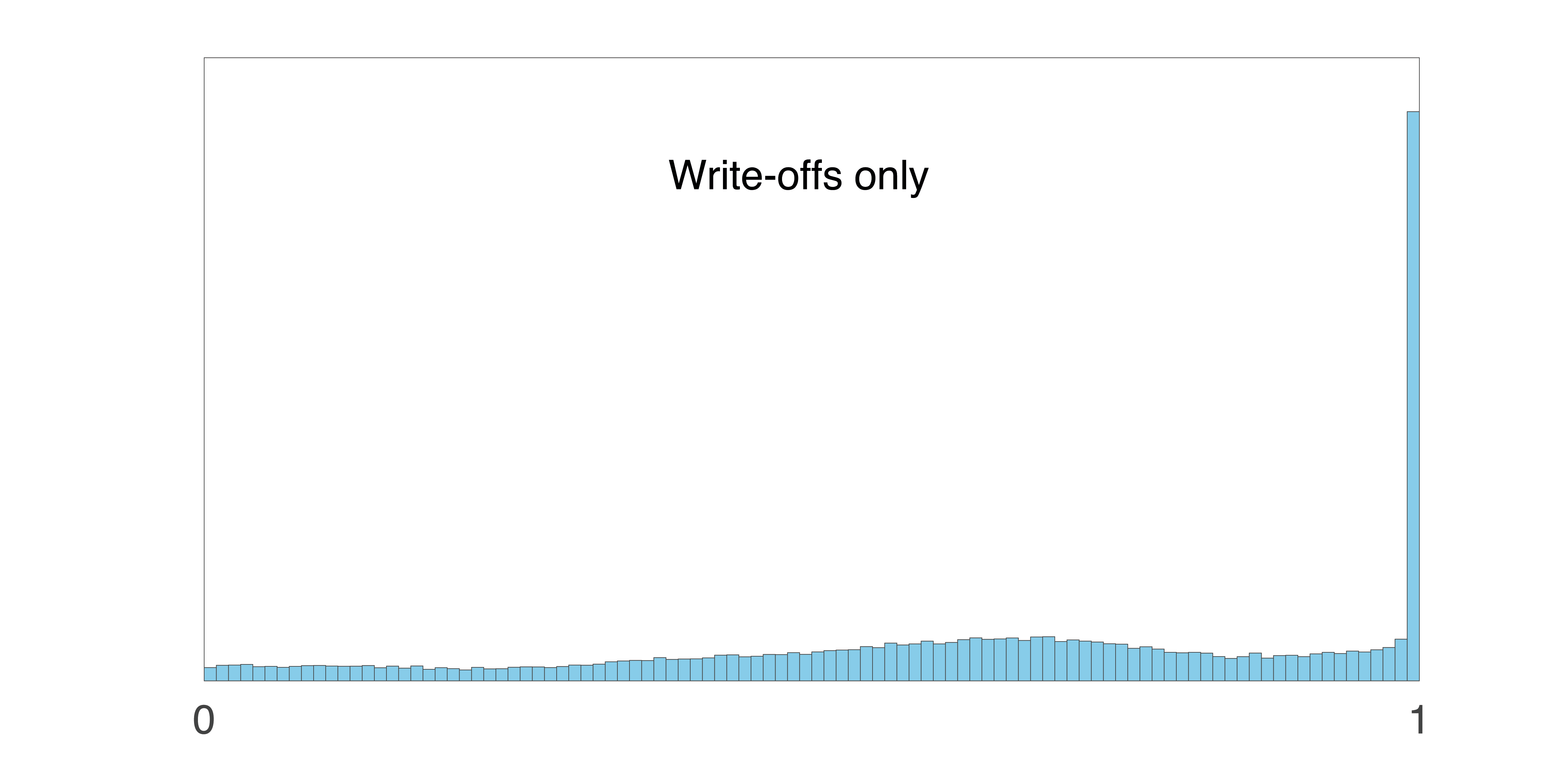}% Inserted image/inset
    }\hspace*{1em}%
  }%
  \caption{Histogram of realised loss rates for the secured ML-portfolio over the full reference period. Graph design follows that of \autoref{fig:PLOutcomes}.}
\label{fig:MLOutcomes}
\end{figure}

The summary statistics for the two portfolios are shown in \autoref{tab:summarystats}. Besides the difference in the average LGD ($\bar{L}$), note the the significant difference in the averages of the workout periods between the two portfolios. Given the lack of collateral, banks typically have stringent write-off criteria for unsecured PLs that shorten the workout process, which contrasts with that of asset-secured MLs.
%agile in recognising a futile recovery process in the case of unsecured loans, and will normally choose to avoid long and expensive workout processes. 
%In contrast, most owner-occupier mortgage borrowers will engage with the bank to arrange a payment holiday or a more affordable repayment schedule when faced with the prospect of losing their home. Even if the property is eventually foreclosed on, the legal and operational processes attached to this action typically take years, rather than months, to resolve. 
Consequently, the average LGD for the PL-portfolio is also much higher than that for the ML-portfolio. The latter's lower average LGD is however accompanied by a higher standard deviation ($\hat{\sigma}$). Interestingly, the means and standard deviations for the PL-portfolio are very similar for the two periods under consideration, whereas the average LGD in respect of ML in the downturn period is somewhat higher at 29.3\% compared to 25.6\% over the full period.
% LGD Graphs
In \autoref{fig:PLOutcomes} and  \autoref{fig:MLOutcomes}, we show the distribution of the aggregated LGD data over the full period, respectively for the PL- and ML-portfolios. As expected, both portfolios have a bimodal shape in the distribution of their LGD-values, though the degree thereof differs by portfolio type. In particular, the ML-portfolio has a substantial proportion of cured accounts that result in zero losses, whereas these cases constitute only a minor proportion for the PL-portfolio. Accordingly, the higher standard deviation of the ML-portfolio is mainly attributed to the high proportion of these cured loans.

\begin{figure}[ht!]
\centering
\includegraphics[width=14.5 cm]{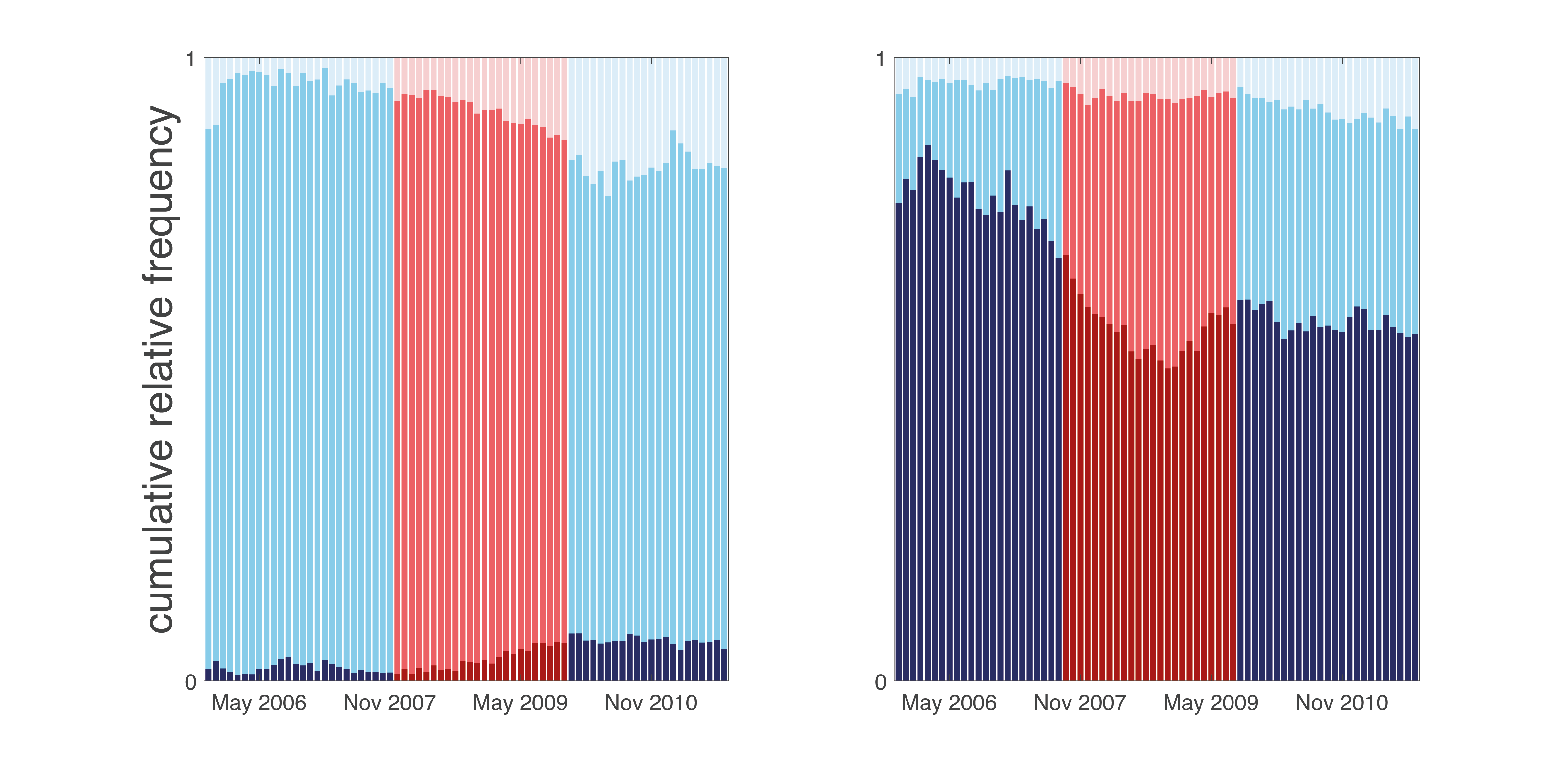}
\caption{Proportion of LGD observations over time at $0$ (dark blue), on the interval $(0,1)$ (medium blue), and at $1$ (light blue). Proportions are shown with respect to PL-portfolio (left panel) and the ML-portfolio (right panel), whilst downturn months are shaded in red.}
\label{fig:Cumulative}
\end{figure}

In \autoref{fig:Cumulative}, we show the proportions of 
LGD-observations over time at various points over the interval $[0,1]$. Evidently, the ML-portfolio experienced a marked drop in the proportion of cured loans during the downturn period, and this proportion did not return to pre-downturn levels before the end of the reference period. A similar pattern emerges in the PL-portfolio, where the proportion of total losses continue to increase well beyond the downturn period, after which it stabilises. This result suggests that the effects of the economic downturn continued well into 2011. The ML-data suggests that the recovery of residential property prices (collateral) lagged behind economic recovery, resulting in greater loan-to-value ratios and fewer defaulted accounts with full recoveries. For both PLs and MLs, the trends could also indicate a change in the bank's underwriting standards. Following the severe losses suffered by banks during the 2008-GFC, many banks adjusted their underwriting standards, as reported by the \citet{FACMinutes} for the US-market. This factor might also be at play in our dataset. 

\section{An illustration of determining \texorpdfstring{$\delta$}{Lg} using the cost of capital approach}
\label{sec:Results}

We first consider a cost of capital of $c=7\%$, as derived from the average cost of equity of approximately $14\%\approx 7\% + 7.2\%$, given a risk-free rate of $r_f=7.2\%$ over 2007-2011; see the financial results from \citet{SBfinreport}. 
We summarise our results in \autoref{tab:resultsforc=678} and note the much higher risk-free rates $r_f$ associated with the macroeconomic downturn period across both PL- and ML-portfolios. This phenomenon clearly reflects the market turbulence that prevailed during the downturn period, which caused the bond yields to surge. %This effect is less pronounced for the ML-portfolio, due to the longer period over which averages are drawn.
Furthermore, we express the EC-amount relative to the portfolio's market-consistent price (MCP) to enable comparison across portfolios and over different terms; a quantity we shall call the EC-ratio. Evidently, this EC-ratio is almost identical between the PL-- and ML-portfolios, despite having very different values for the average loss $\bar{L}$. The significantly longer workout period of the ML-portfolio explains this fact, with an average time span (25.64 months) that is more than five times longer than that of the PL-portfolio (4.67 months). Another factor is the greater standard deviation $\hat{\sigma}$ of $L$ for the ML-portfolio with its smaller losses, and vice versa for the PL-portfolio.
The greater average losses $\bar{L}$ of the PL-portfolio also explains its greater discount rates $r_d$ relative to that of the ML-portfolio, which implies that the risk margin $R$ affects $r_d$ more stringently than the recoveries.

%Despite the very similar EC amounts, and the lower average risk-free rate, $r_f$,  the discount rate, $r_d$, is considerably higher for the PL-portfolio than for the ML-portfolio over the full period. This is attributable to the much lower average recovery rate (1 - $\bar{L}$) for PL's, resulting in $R$ carrying a significantly higher weight than the recoveries in the determination of $r_d$. 

When comparing the downturn with the full period, the greater EC-ratio (1.653 vs 1.469) of the ML-portfolio is largely due to the portfolio's longer average workout period (33.59 vs 25.64 months) and its greater $\bar{L}$-value (29\% vs 25.6\%). However, this effect is offset by the greater $r_f$-value, which results in risk premiums $\delta$ that are very similar (5.4\% vs 5.5\%) between the downturn and full period.
%When comparing the ML-portfolio over the downturn period with that over the full period, the EC as a proportion of $P$ is considerably higher (1.653 vs 1.469). This is attributable to both the higher $\bar{L}$ (29\% vs 25.6\%) and the longer average recovery period (33.59 vs. 25.64 months). The lower recovery rate and higher EC amount result in a higher discount rate, $r_d$. However, because of the higher risk-free rate, $r_f$, over the downturn period, the risk premiums are in fact very similar (5.4\% vs. 5.5\%).
The difference in $r_d$ (15.21\% vs 17.06\%) is larger for the PL-portfolio, which is almost entirely attributable to the greater difference in $r_f$-values (8.93\% vs 6.37\%). Greater $r_f$-values also explain why the PL-portfolio's resulting $\delta$-value is lower during the downturn than the full period, which is otherwise curious. These results however corroborate that of \citet{scheule2020benchmarking} in that lower $\delta$-values are associated with periods of greater $r_f$-values, especially when applying the ME-variant of the ROE-method in determining $r_d$.
%For the PL-portfolio it is notable that both the variance and the EC amount is slightly higher for the full period than for the downturn period. Considering that the full period encompasses the downturn period, and then also include divergent (prosperous) months with much lower losses, this result seems sensible. When comparing the PL discount rates, it is evident that the downturn period produces a much higher rate than the full period, despite fairly similar values for $\bar{L}$ and $\hat{\sigma}^2$. This can almost entirely be ascribed to the higher risk-free rate, $r_f$, that applied over the downturn period. The risk premium, $\delta$, for PL is slightly lower than that for the full period due to the considerably higher risk-free rate, over that period. For ML the downturn risk premium is only marginally higher than for the full period, due to the longer recovery term. This result is consistent with that reported by \cite{scheule2020benchmarking}. The authors find lower risk premiums associated with periods of higher risk-free rates when applying a variety of market equilibrium methods to determine the LGD discount rate risk premium. 

\begin{table}[hbt!]
\centering
\caption{Comparing various metrics across the PL-- and ML-portfolios over two periods of time, having set the cost of capital rate at $c=\{6\%,7\%,8\%\}$. Metrics include the calculated risk-free rate $r_f$, the associated discount rate $r_d$, the resulting risk premium $\delta$, the EC-ratio $\mathrm{EC}/\mathrm{MCP}$, the average loss $\bar{L}$, and the sample standard deviation of losses $\hat{\sigma}$. Boldfaced values indicate the baseline scenario of $c=7\%$.} \label{tab:resultsforc=678}
\begin{threeparttable}
%\label{table_example}
\begin{tabular}{lccccccc}
	\toprule
	& $c$ & $r_f$ & $r_d$ & $\delta$ & EC/$\mathrm{MCP}$ & $\bar{L}$ & $\hat{\sigma}$\\
 \midrule
\multicolumn{8}{c}{Personal Loans} \\
\midrule
\multirow{3}{*}{Full period} & $6\%$& $6.37\%$& $13.90\%$& $7.53\%$& $1.441$& $0.747$& $0.293$\\
& \textbf{7\%} & \textbf{6.37\%} & \textbf{15.21\%} & \textbf{8.84\%} & \textbf{1.449} & \textbf{0.749} & \textbf{0.290}\\
& $8\%$& $6.37\%$& $16.52\%$& $10.15\%$& $1.455$& $0.751$& $0.288$\\
\midrule
\multirow{3}{*}{Downturn period} & $6\%$& $8.93\%$& $15.86\%$& $6.93\%$& $1.424$& $0.758$& $0.275$\\
& \textbf{7\%} & \textbf{8.93\%} & \textbf{17.06\%} & \textbf{8.13\%} & \textbf{1.430} & \textbf{0.760} & \textbf{0.273} \\
& $8\%$& $8.93\%$& $18.27\%$& $9.33\%$& $1.436$& $0.762$& $0.270$\\
\midrule
\multicolumn{8}{c}{Mortgage Loans}\\
\midrule
\multirow{3}{*}{Full period} & $6\%$& $6.88\%$& $11.45\%$& $4.57\%$& $1.445$& $0.253$& $0.364$\\
& \textbf{7\%} & \textbf{6.88\%} & \textbf{12.28\%} & \textbf{5.40\%} & \textbf{1.468} & \textbf{0.256} & \textbf{0.365} \\
& $8\%$& $6.88\%$& $13.14\%$& $6.26\%$& $1.493$& $0.258$& $0.367$\\
\midrule
\multirow{3}{*}{Downturn period} & $6\%$& $8.49\%$& $13.13\%$& $4.64\%$& $1.624$& $0.288$& $0.369$\\
& \textbf{7\%} & \textbf{8.49\%} & \textbf{13.99\%} & \textbf{5.50\%} & \textbf{1.653} & \textbf{0.290} & \textbf{0.370} \\
& $8\%$& $8.49\%$& $14.86\%$& $6.37\%$& $1.684$& $0.293$& $0.372$\\
\bottomrule
\end{tabular}
\end{threeparttable}
\end{table}

We highlight the following two interesting results in \autoref{tab:resultsforc=678}. 
Firstly, note the $r_d$-rates across the downturn and full periods, which are respectively 13.99\% and 12.28\% for the ML-portfolio, and 17.06\% and 15.21\% for the PL-portfolio. The range of these $r_d$-values align with those obtained by \citet{scheule2020benchmarking}, who analysed corporate loan recovery data from South Africa. Their estimated ROE-based $r_d$-values range between 12\% and 17\% over the full period, where 17\% coincided with the downturn period. Their average cost of equity of 14\% also compares well with our chosen value of 14\% thereof.
Moreover, their calculated equilibrium rates (or the ME-variant of ROE within our context) range between 8\% and 12.5\%, and it peaks during mid-2008. The WACC-based $r_d$-values follow the same pattern, but are approximately 50 basis points lower than the equilibrium rates over the same period. 
As for our second result, consider the cost of capital in \autoref{tab:resultsforc=678} and compare its values across portfolios: a larger $c$-value implies a greater $r_d$-value. Moreover, it appears that $c$-values have a much more pronounced effect on the $r_d$ of the PL-portfolio than that of the ML-portfolio. This phenomenon is largely explained by the greater $\bar{L}$-values associated with the unsecured PL-portfolio. In this case, the present value of the recoveries has a relatively lower weight in determining the $r_d$-value than what would have been the case with the ML-portfolio.
In particular, a 1\% increase in $c$ results in an approximate increase of 1.2\%-1.3\% in $r_d$ for the PL-portfolio, versus a 0.83\%-0.88\% increase in $r_d$ for the ML-portfolio.

%It can be seen from \autoref{eq:RM} and \autoref{eq:objF} that besides the risk margin, $R$, there are two further drivers of the MCP, and therefore the discount rate, $r_d$. These are the expected recoveries, $Y$, and the risk-free rate, $r_f$. $R$, in turn, has $c$, as well as $\bar{L}$ and $\hat{\sigma}$, as drivers.

\section{Conclusion}
\label{sec:Conclusion}

We have demonstrated how to infer a market-consistent discount rate $r_d$ for LGD-modelling purposes by calculating a \textit{market consistent price} (MCP) for a portfolio of defaulted debt. Determining such an MCP relies on the cost of capital that is required for holding defaulted debt in the first place, as inspired by the approach followed within the Solvency II regulatory regime. In our method, we calculated the \textit{economic capital} (EC) for post-default loss risk by using the single factor model proposed by \citet{tasche2004}. However, we anticipate that banks will likely use their own internal capital models for this purpose. Given an EC-amount, we presented an algorithm for iteratively calculating the MCP as the difference between the cost of holding EC (or risk margin) and the discounted recoveries. In so doing, the risk premium $\delta$ is ultimately calculated within each iteration towards deriving a final $r_d$-value.

We found that the main drivers of $r_d$ include the following: 1) the mean and variation of recovery cash flows; 2) the average risk free rate $r_f$ that applies over the reference period; 3) the average term of the cash flows during the workout period; and 4) the cost of capital rate. As expected, greater volatility in the recovery cash flows is associated with greater risk, and hence, a higher $r_d$-value. The same holds true for longer recovery terms. 
Similar to the findings of \citet{scheule2020benchmarking}, we found that a greater $r_f$-value is, ceteris paribus, associated with a lower risk premium $\delta$. This result renders the overall $r_d$-value as surprisingly stable over multiple types of time periods, e.g., downturn vs full macroeconomic period. 
Furthermore, $r_d$ is fairly sensitive to the cost of capital rate, and even more so when recovery rates are low. This sensitivity seems sensible when considering the cost of raising and servicing risk capital in guarding against variability in post default recoveries. At the same time, consider that the cost of equity (in excess of the risk-free rate) should reflect a bank's overall riskiness and its cost of funding, which further justifies the strong link between $r_d$ and the cost of capital.

% Future work
As for future work, one might compare our resulting LGD-values discounted by $r_d$ to those discounted by the ever-popular average contract rate, particularly since our dataset prohibited such a comparison. It may also be useful to compare the two resulting LGD-models that are built from these disparate LGD-values, which are discounted differently.
Ultimately, our proposed method satisfies the regulatory requirement in that $r_d$ should account for both the time value of money, as well as include a risk premium for undiversifiable risk in the recovery cash flows. This method can be easily applied throughout downturn and normal `through-the-cycle' (TTC) macroeconomic conditions. We therefore deem our method appropriate in producing a market-consistent $r_d$-value towards estimating the LGD-component in both regulatory and economic capital modelling.
%From \citet{scheule2020benchmarking}, the contract rate remains the most popular for use in practice, presumably in part due to its prescription for provisioning purposes by the IFRS Accounting Standard from the \citet{ifrs9_2014}. Due to data limitations, a comparative study between the proposed CoC-discount rate and the average contract rate on the ML and PL portfolios was excluded from the scope of our study. To provide a more comprehensive analysis of the CoC-discount rate, future research could consider such a comparative study to fully explore the difference between the CoC-risk premium and that embedded in the loan contract rate.

%The proposed methodology satisfies the regulatory requirement for the LGD discount rate to account for the time value of money and to include a risk premium that reflects the undiversifiable risk of recovery cash flows. The methodology can easily be applied to through-the-cycle and downturn conditions and, as such, could be regarded as appropriate for the determination of an LGD discount rate for both regulatory and economic capital purposes.

%TC:ignore

%\appendix
%\input{6-Appendix}

%--------------------------------------------------------%
%	REFERENCE LIST
%--------------------------------------------------------%

\singlespacing
\printbibliography % using biblatex
%\section*{References}
%\bibliographystyle{newapa}
% see http://texdoc.net/texmf-dist/doc/latex/natbib/natbib.pdf for more styles
%\bibliography{bibliography} 

\onehalfspacing

%TC:endignore

%--------------------------------------------------------%
%	END OF DOCUMENT
%--------------------------------------------------------%

\end{document}